\documentclass[runningheads]{llncs}

\usepackage[english]{babel}

\usepackage{graphicx}

\usepackage{url}

\usepackage{amsmath}
\usepackage{amssymb}
\usepackage{color}

\usepackage{xhfill}
\usepackage{longtable}
\usepackage{makecell}
\usepackage{booktabs}
\usepackage{multirow}
\usepackage{enumerate}
\usepackage[shortlabels]{enumitem}
\setlist{nosep}

\usepackage[htt]{hyphenat}
\usepackage{forest}
\usepackage[skins, breakable]{tcolorbox}

\usepackage{rotating}

\usepackage{algorithm}
\usepackage{algpseudocode}
\usepackage{listings}

\usepackage{tikz}

\usepackage[hidelinks]{hyperref}

\usepackage{cleveref}
\crefformat{footnote}{#2\footnotemark[#1]#3}
\usepackage[bottom]{footmisc}

\usepackage{stfloats}

\newcommand{\etal}{\textit{et al.}}

\newcommand{\q}[1]{``#1''}

\newcolumntype{H}[1]{>{\centering\arraybackslash}p{#1\textheight}}
\newcolumntype{h}[1]{>{\arraybackslash}p{#1\textheight}}
\newcolumntype{V}[1]{>{\centering\arraybackslash}p{#1\textwidth}}
\newcolumntype{v}[1]{>{\arraybackslash}p{#1\textwidth}}

\lstdefinelanguage{json}{
    basicstyle=\small\ttfamily,
    numbers=left,
    numberstyle=\scriptsize,
    stepnumber=1,
    numbersep=8pt,
    showstringspaces=false,
    breaklines=true,
    frame=lines,
    backgroundcolor=\color{white},
}

\begin{document}

\title{Investigating the Effectiveness of Bayesian Spam Filters in Detecting LLM-modified Spam Mails}

%
\titlerunning{Effectiveness of Bayesian Spam Filters against LLM-modified Spam Mails}
%
\author{Malte Josten\orcidID{0000-0003-2102-1575},
Torben Weis\orcidID{0000-0001-6594-326X}}
\authorrunning{M. Josten and T. Weis}
%
\institute{University of Duisburg-Essen\\Duisburg, Germany\\
\email{\{malte.josten,torben.weis\}@uni-due.de}}

\maketitle              

\begin{abstract}
Spam and phishing remain critical threats in cybersecurity, responsible for nearly 90\% of security incidents.
As these attacks grow in sophistication, the need for robust defensive mechanisms intensifies.
Bayesian spam filters, like the widely adopted open-source SpamAssassin, are essential tools in this fight.
However, the emergence of large language models (LLMs) such as ChatGPT presents new challenges.
These models are not only powerful and accessible, but also inexpensive to use, raising concerns about their misuse in crafting sophisticated spam emails that evade traditional spam filters.
This work aims to evaluate the robustness and effectiveness of SpamAssassin against LLM-modified email content.
We developed a pipeline to test this vulnerability.
Our pipeline modifies spam emails using GPT-3.5 Turbo and assesses SpamAssassin's ability to classify these modified emails correctly.
The results show that SpamAssassin misclassified up to 73.7\% of LLM-modified spam emails as legitimate.
In contrast, a simpler dictionary-replacement attack showed a maximum success rate of only 0.4\%.
These findings highlight the significant threat posed by LLM-modified spam, especially given the cost-efficiency of such attacks (0.17 cents per email).
This paper provides crucial insights into the vulnerabilities of current spam filters and the need for continuous improvement in cybersecurity measures.

\end{abstract}

\section{Introduction}\label{sec:intro}
Spam and phishing are persistent threats, accounting for nearly 90\% of security incidents \cite{cisa90}.
The prevalence and sophistication of these attacks necessitate continuous advancements in defensive mechanisms.
Among the various tools employed to combat spam, Bayesian spam filters have been widely adopted \cite{assp}\cite{bogofilter}\cite{dspam}\cite{spambayes}\cite{sa-why-rules}.
In this work, we only evaluate \textit{SpamAssassin}\footnote{\url{https://spamassassin.apache.org}}, as it is an open-source software that offers transparency and allows for community-driven enhancements, ensuring continuous improvement and security updates.
However, the advent of large language models (LLMs) introduces new challenges and opportunities in the cybersecurity landscape.
LLMs are omnipresent, easily accessible, inexpensive to use, and exhibit remarkable power in generating and modifying text.
This raises concerns about their potential misuse in reformulating (rephrasing) existing spam emails that can evade traditional spam filters.
To address this emerging threat, we have developed a comprehensive pipeline to evaluate the robustness and effectiveness of Bayesian spam filters, specifically targeting SpamAssassin and its performance against LLM-modified email contents.
Our research aims to illuminate the capabilities of LLMs in bypassing these filters and to identify potential vulnerabilities that need to be addressed.
The findings underscore the importance of enhancing spam filters to counteract the sophisticated manipulations enabled by modern language models.
In this paper, we present the results of our experiments, providing insights into our developed pipeline and demonstrate the (in)effectiveness of one of the most-used open-source spam filters, SpamAssassin, against LLM-modified spam emails.

\section{Related Work}\label{sec:related-work}

In their article, Gallagher \etal{} \cite{gallagherPhishingSocialEngineering2024} discuss the evolving threats posed by LLMs in cybercrime, particularly in phishing and social engineering.
They observe that the traditional high-volume, low-quality spamming techniques have increasingly been replaced by more sophisticated, targeted spear phishing methods that require greater effort.
Our approach significantly reduces this effort; while it is fully automated for tasks other than the actual sending of spam emails and is also cost-effective, it is not specifically tailored for spear phishing attacks.
Moreover, the authors highlight that despite existing guidelines, LLMs remain susceptible to manipulation through prompt engineering, a challenge detailed further in \Cref{sec:eval} of this paper.

Motlagh \etal{} also explore the adversarial and defensive use cases for LLMs \cite{motlaghLargeLanguageModels2024}.
The authors outline the various ways in which LLMs are already misused, such as gathering information to prepare an attack (reconnaissance), gaining unauthorized access to target systems, evading detection, and creating malicious code.
They also include research by Karanjai \cite{Karanjai-2022}, who already investigated the ability of various LLMs to generate convincing phishing emails.
However, Karanjai's work did not specify a proper evaluation metric to measure the effectiveness of these emails.
In contrast, we aim to refine and reformulate existing spam emails that are currently unable to pass the filter-under-test (FUT).
We further provide two metrics in the form of the success rate and semantic similarity of the LLM-modified texts to analyze the resulting email contents.

As part of their research, Webb \etal{} \cite{webbExperimentalEvaluationSpam2005a} identified the potential risk of \textit{camouflaged} spam emails, which blend elements of both spam and legitimate content, complicating detection by conventional spam filters.
They show that filters that were originally only trained on solely ham and spam emails struggle to deal with camouflaged emails, with the Naive Bayes filter performing particularly poorly.
Motivated by these findings, our study seeks to delve deeper into the limitations of Bayesian filters.
Unlike Webb \etal{}, we do not focus on emails explicitly crafted from mixed spam and ham content.
Instead, we aim to carefully rephrase certain parts of emails to investigate potential vulnerabilities in Bayesian spam filters.

\section{Experimental Setup}\label{sec:exp-setup}
Our pipeline\footnote{\url{https://github.com/MalteJosten/llm-spam-test}} is designed to target a mail server configuration that uses a spam filter and offers an API for accessing and modifying data on received emails.
Specifically, our focus is on SpamAssassin, which serves to classify incoming emails as either legitimate (ham) or spam using a Bayesian classifier.
Typically, SpamAssassin is set up as a gateway, filtering emails before they reach the inbox and blocking those identified as spam.
However, in our setup, it functions differently by assigning a spam score to emails post-receipt.


\begin{figure}[!b]
    \centering
    \resizebox{.95\textwidth}{!}{
        \begin{tikzpicture}
            \node[draw, fit={(0,0) (1.75,1)}, inner sep=0pt, label=center:LLM] (llm) {};
            \node[draw, fit={(1.75,0) (2.25,1)}, inner sep=0pt, label={[rotate=90]center:API}] (llm-api) {};
            \node[draw, fit={(3.75,0) (6,1)}, inner sep=0pt, label={[align=center]center:Pipeline}] (pipeline) {};
            \node[draw, fit={(7.5,0) (8,1)}, inner sep=0pt, label={[rotate=90]center:API}] (mailpit-api) {};
            \node[draw, fit={(8,0) (9.75,1)}, inner sep=0pt, label=center:Mailpit] (mailpit) {};
            \node[draw, fit={(7.5,-0.75) (9.75,-1.75)}, inner sep=0pt, label=center:SpamAssassin] (spamassassin) {};
    
            \draw[<->] (llm-api.east) -- node[below,midway] {(a)} (pipeline.west);
            \draw[->] (pipeline.north) -- (4.875,1.25) -- node[above,midway] {(b)} (8.875,1.25) -- (mailpit.north);
            \draw[<->] (mailpit.south) -- node[left,midway] {(c)} (8.875,-0.75);
            \draw[<->] (pipeline.east) -- node[below,midway] {(d)} (mailpit-api.west);
        \end{tikzpicture}
    }
    \caption{To rephrase a spam email and test it's resulting classification, the pipeline communicates with various components to (a) modify the spam email body, (b) send the email via SMTP to the mail server, and (d) retrieve the classification label through the Mailpit API. With the help of SpamAssasin, (c) Mailpit labels the email as either ham or spam.}
    \label{fig:sys-model}
\end{figure}
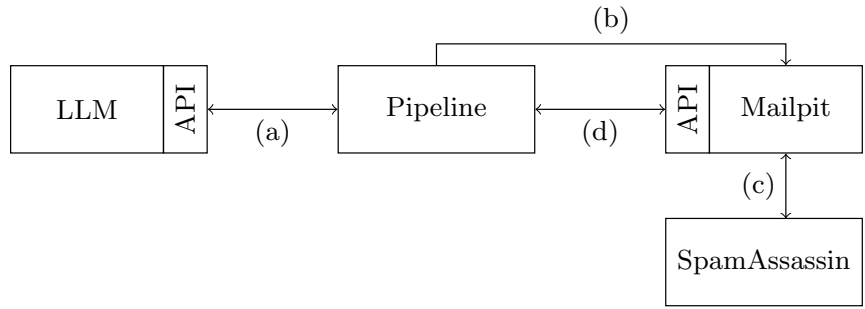

\Cref{fig:sys-model} depicts the system architecture that demonstrates both, the relations between the necessary components, and the data flow.
For simplicity's sake, the system-under-test (SUT) comprises two docker images: one containing a mail server, realized through Mailpit \cite{axllent:mailpit} and the other image provides a SpamAssassin instance that gets integrated into the former \cite{axllent:spamassassin}.
Internally, the SpamAssassin image utilizes Linux's \texttt{spamassassin} service\footnote{\url{https://linux.die.net/man/1/spamassassin}} with its default configuration.

To rephrase a spam email and to test its resulting classification and the overall robustness of the FUT, our pipeline communicates with various components.
The core part, having access to the LLM's API, facilitates the possibility of reformulating and modifying an existing spam email, intending to make it sound less aggressive and spam-like.
Thus, the modified emails acquired through (a) are sent via SMTP to the mail server and subsequently classified by SpamAssassin (c).
By utilizing Mailpit's API (d), we retrieve the spam classification for the email and use it to evaluate our method's Success Rate as well as the effectiveness of the FUT against LLM-modified spam emails.

\section{Method}\label{sec:model}
\begin{figure}[!b]
    \centering
        \begin{tikzpicture}
            \node[circle, fit={(-1.5,0.25) (-1,0.75)}, fill=black,inner sep=0pt] (start) {};
            \node[draw, thick, fit={(0,0) (2.25,1)}, rounded corners=.2cm, inner sep=0pt, label={[align=center]center:1. Pre-\\Processing}] (pp) {};
            \node[draw, thick, fit={(3.25,0) (5.5,1)}, rounded corners=.2cm, inner sep=0pt, label={[align=center]center:2. LLM\\Modification}] (llm) {};
            \node[draw, thick, fit={(6.5,0) (8.75,1)}, rounded corners=.2cm, inner sep=0pt, label={[align=center]center:3. Robustness\\Evaluation}] (eval) {};

            \draw[->, thick] (start.east) -- (pp.west);
            \draw[->, thick] (pp.east) -- (llm.west);
            \draw[->, thick] (llm.east) -- (eval.west);

            \node[draw, fit={(-.125,-0.5) (2.375,-1.25)}, inner sep=0pt, label={[align=center]center:Spam Dataset}] (dataset) {};
            \node[draw, fit={(3.25,1.5) (5.5,2.5)}, inner sep=0pt, label={[align=center]center:Mailpit/\\SpamAssassin}] (mail) {};
            \node[draw, fit={(3.25,-0.5) (5.5,-1.25)}, inner sep=0pt, label={[align=center]center:OpenAI API}] (openai) {};

            \draw[->, dashed, thick] (dataset.north) -- (pp.south);
            \draw[<->, dashed, thick] (openai.north) -- (llm.south);
            \draw[<->, dashed, thick] (pp.north) -- (1.125,2) -- (mail.west);
            \draw[<->, dashed, thick] (eval.north) -- (7.65,2) -- (mail.east);
        \end{tikzpicture}
    \caption{Pipeline to (1) pre-process spam emails extracted from a spam dataset, (2) hand them over to the OpenAI API to be modified by the LLM, and (3) evaluate the spam filters robustness against the rephrased email bodies.}
    \label{fig:pipeline}
\end{figure}
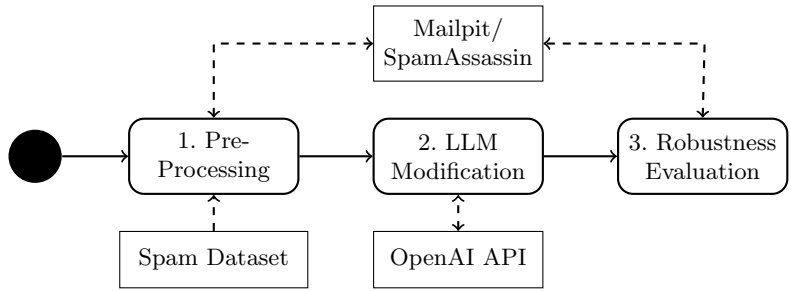

We designed and implemented a pipeline used to test the efficacy of LLM-modified spam email contents.
This pipeline can be split up into three major parts: (1) pre-processing, (2) LLM modification, and (3) robustness evaluation.
\Cref{fig:pipeline} depicts these parts and puts them into context of the other system components.

\subsection{Pre-processing}
For this setup, we use the spam-labeled emails from the SpamAssassin Public Mail Corpus\footnote{\url{https://spamassassin.apache.org/old/publiccorpus/}} dataset (2.399 spam emails in total).
First, we convert all emails to \texttt{.eml} files to ensure they can be used more efficiently by the SMTP python library.
Next, the STMP header fields \texttt{Bcc}, \texttt{Cc}, \texttt{From}, and \texttt{To} were anonymized (for example to \textit{from@example.com}).
This eliminates any speculations regarding the addresses' impact during the classification.
We further chose to work with two datasets \textit{original} and \textit{minimal}, as SMTP headers are usually considered during the classification process and thus contribute to the decision of whether an email is considered legitimate or not \cite{Khater-2012}\cite{Gascon-2018}\cite{Abawajy-2013}\cite{Nabeel-2021}.
This decision is also supported by the rise in misclassifications of spam mails after removing the headers (see \Cref{tbl:pre-proc})
It is also important to note that we update the \texttt{Date} field to the current date and time to further minimize the SMTP header's impact on the classification.
\vspace*{2mm}
\begin{description}
    \item[Original] uses all the original SMTP headers \q{as they are} (except the anonymi-zation and the updated \texttt{Date} field).
    \item[Minimal] only contains the mandatory SMTP header fields \texttt{Date} and \texttt{From} (as defined by RFC 2822 \cite{rfc2822}).
        In addition, we also include the syntactically optional \texttt{Bcc}, \texttt{Cc}, \texttt{Subject}, and \texttt{To} fields, to make the dataset more realistic.
\end{description}
\vspace*{2mm}
Working with the \textit{minimal} dataset ensures that the SMTP body's content is the main factor during the spam classification process.
Hence, we can reason that any changes in the results can clearly be attributed to the SMTP body.
Additionally, we can use \textit{original} to compare our findings.

In the course of the dataset preparation (\textit{Raw} $\rightarrow$ \textit{Post-Mod}), some emails are filtered out due to, for example, encoding or formatting errors.
After cleaning and updating the datasets, we define the ground truth, which consists of emails labeled as spam by both the dataset and the spam filter.
Emails not detected by the spam filter are excluded.
This ensures a more accurate and reliable evaluation of the success rate.
We send each of these emails to the Mailpit instance (\textit{Sent}) and retrieve their spam classification.
Based on the gathered classifications, we build the ground truth, only using the correctly classified spam emails.
Hence, for our work, we only consider 1.457 for the \textit{original}, and 189 spam emails for the \textit{minimal} dataset (\textit{Spam}), as they constitute the ground truth.

\begin{table}[htp]
    \centering
    \caption{Number of emails during the first step: 1. Pre-processing.}
    \label{tbl:pre-proc}
   \begin{tabular}{l|ll|lll|r}
        \toprule
        \multirow{2}{*}{\textbf{Dataset}} & \multirow{2}{*}{\textbf{Raw}} & \multirow{2}{*}{\textbf{Post-Mod}} & \multirow{2}{*}{\textbf{Sent}} & \multicolumn{2}{c|}{\textbf{Classification}} & \multirow{2}{*}{\textbf{\makecell[l]{Misclass. Rate}}} \\
        & & & & \textbf{Ham} & \textbf{Spam} & \\
        \midrule
        \textit{original} & 2.399 & 2.399 & 2.361 & 904   & 1.457 & 38.3\% \\
        \textit{minimal}  & 2.399 & 2.111 & 2.091 & 1.902 & 189   & 91\% \\
        \bottomrule
    \end{tabular}
\end{table}

\subsection{LLM Modification}
To test the spam filter’s robustness against LLM-modified content, we employed the OpenAI API with GPT-3.5 Turbo as the model to rephrase the email bodies.
Prompts \ref{prompt:P1} and \ref{prompt:P2} were used in conjunction with the '\texttt{system}' role to prime the model by providing it some context in the form of the SMTP body, as well as instructions on how to answer and what (not) to include in the response \cite{openai:text-gen}.
Prompt \ref{prompt:P3} is the actual user prompt that triggers the LLM's response generation.

\vspace*{2mm}
\begin{enumerate}[label=\textbf{(P\arabic*)}, ref=P\arabic*, leftmargin=2\parindent]
    \item \label{prompt:P1} Keep in mind the following text I wrote: $\ll$ SMTP body $\gg$
    \item \label{prompt:P2} Give your answer as a JSON object, only using the fields defined by the \texttt{print\_result} function.
        I only need the response, no additional text.
        Don't generate a subject line.
        Instead of using placeholders, just leave out the placeholder brackets.
        You're allowed to use line breaks in your answer.
        Preserve all links.
    \item \label{prompt:P3} Please rephrase the previous content to be less aggressive and replace spam-like words and formulations.
\end{enumerate}
\vspace*{2mm}

Since simple text instructions did not yield consistent responses, thus preventing automated processing, our system utilizes OpenAI's function calling capability to make and retrieve the results of the API request.
Inspired by the work of Koide \etal{}, we established the function call \texttt{print\_result} \cite{Koide-2024}.
The \texttt{is\_success} property indicates whether the LLM was able to comply with the request or not and implies, if, for example, the request violates the model's ethical guidelines.
If the request failed, the user gets an explanation of why it failed (\texttt{failed-description}) and a keyword, summarizing the reason for failing (\texttt{failed-keyword}), respectively.
In \Cref{sec:eval}, the latter is utilized to show the shortcomings of this prompt engineering approach, which in turn may help to understand the model's restrictions and enable further refinement of the function call.

\begin{table}[!b]
    \centering
    \caption{Properties of functional call \texttt{print\_result}}
    \begin{tabular}{v{.24}v{.075}v{.55}}
        \toprule
        \textbf{Property} & \textbf{Type} & \textbf{Description} \\
        \midrule
        \texttt{is\_success}        & boolean & A boolean value indicating whether the text could successfully be reformulated (true) or whether reformulating the text was not possible due to your rules (false). \\
        \texttt{failed-description} & string  & Detailed description on why the text could not be reformulated. \\
        \texttt{failed-keyword}     & string  & One keyword summarizing the main reason why the text could not be reformulated. \\
        \texttt{body}               & string  & If \texttt{is\_success} is true, this field is filled with the reformulated text. Otherwise, this field is empty. \\
        \bottomrule
    \end{tabular}
    \label{tbl:fc-properties}
\end{table}

Importantly, only the email body was modified, with no changes made to the headers by the LLM.
The modified SMTP body was then merged with the SMTP headers to create new spam emails with altered content but consistent metadata.

\subsection{Robustness Evaluation}
To assess the effectiveness of the modified spam emails in evading the spam filter, the newly constructed spam emails were sent to the mail server for classification by SpamAssassin.
Using the mail server API, we extracted the spam ratings and documented the classification results.
Consequently, we were able to then calculate the success rate by determining how many of the modified spam emails were misclassified as legitimate.
Additionally, we analyzed the extent to which the content's semantics had changed to understand the impact of LLM modifications on the emails' effectiveness.
The following section presents the results of both, the success rate and the semantic similarities, respectively.

\section{Evaluation}\label{sec:eval}
In the course of this evaluation, we want to look at the success rate of our approach (also in comparison to a rather simple dictionary-replacement attack approach), what might be a factor in hindering the success rate to be even higher, and whether our approach is able to preserve some form of semantic similarity between the original and the LLM-modified texts.

\begin{table}[!b]
    \centering
    \caption{Number of emails during the second step: 2. LLM-modification.}
    \label{tbl:llm-mod}
    \begin{tabular}{v{.167}|v{.09}v{.125}v{.12}v{.14}}
        \toprule
        \textbf{Dataset} & \textbf{Initial} & \textbf{Pre-LLM} & \textbf{Rejected} & \textbf{Post-LLM} \\
        \midrule
        \textit{original} & 1.457 & 1.401 & 262 & 1.139 \\
        \textit{minimal}  & 189   & 181   & 44  & 137 \\
        \bottomrule
    \end{tabular}
\end{table}
\begin{table}[!b]
    \centering
    \caption{Number of emails during the third step: 3. Robustness Evaluation.}
    \label{tbl:success}
    \begin{tabular}{v{.167}|v{.06}v{.06}|v{.05}v{.05}|v{.175}}
        \toprule
        \multirow{2}{*}{\textbf{Dataset}} & \multirow{2}{*}{\textbf{Initial}} & \multirow{2}{*}{\textbf{Sent}} & \multicolumn{2}{c|}{\textbf{Classification}} & \multirow{2}{*}{\textbf{Success Rate}} \\
        & & & \textbf{Ham} & \textbf{Spam} & \\
        \midrule
        Dictionary & & & & & \\
        \makecell[r]{\textit{original}} & 1.457 & 1.457 & 6 & 1.451 & \makecell[r]{0.4\%} \\
        \makecell[r]{\textit{minimal}}  & 189 & 189 & 0 & 189 & \makecell[r]{0\%} \\
        \midrule
        LLM-modified & & & & & \\
        \makecell[r]{\textit{original}} & 1.139 & 1.135 & 107 & 1.028 & \makecell[r]{9.4\%} \\
        \makecell[r]{\textit{minimal}}  & 137 & 137 & 101 & 36 & \makecell[r]{73.7\%} \\
        \bottomrule
    \end{tabular}
\end{table}

The effectiveness of the proposed approach is primarily assessed by measuring the rate at which modified spam emails are incorrectly labeled as legitimate emails.
This metric of success is quantified using the success rate, which is defined as the ratio of emails classified as ham to the number of emails sent, as shown in \Cref{tbl:success}.
Employing our pipeline and applying our evaluation metric, we can see that 9.4\% (\textit{original}) and 73.7\% (\textit{minimal}) of the modified spam emails that were previously labeled as spam (\textit{Spam} in \Cref{tbl:pre-proc}) are now misclassified as ham.
One can observe an even more remarkable success rate of 95.8\% when considering the initial number of spam emails (\textit{Sent} in \Cref{tbl:pre-proc}) in relation to emails falsely labeled as ham (\textit{Ham} in \Cref{tbl:pre-proc} and \Cref{tbl:success}).
However, it remains unclear whether these results can be entirely attributed to our pipeline or if the obsolescence of the dataset (almost 20 years old) and the choice of FUT are the driving factors.
Based on the noticeable improvement from removing the non-required SMTP header fields (\textit{original} $\rightarrow$ \textit{minimal}), we can conclude that rephrasing the email's body actually has a great impact on its (mis)classification.

\begin{figure}[!b]
    \centering
    \includegraphics[width=.85\textwidth]{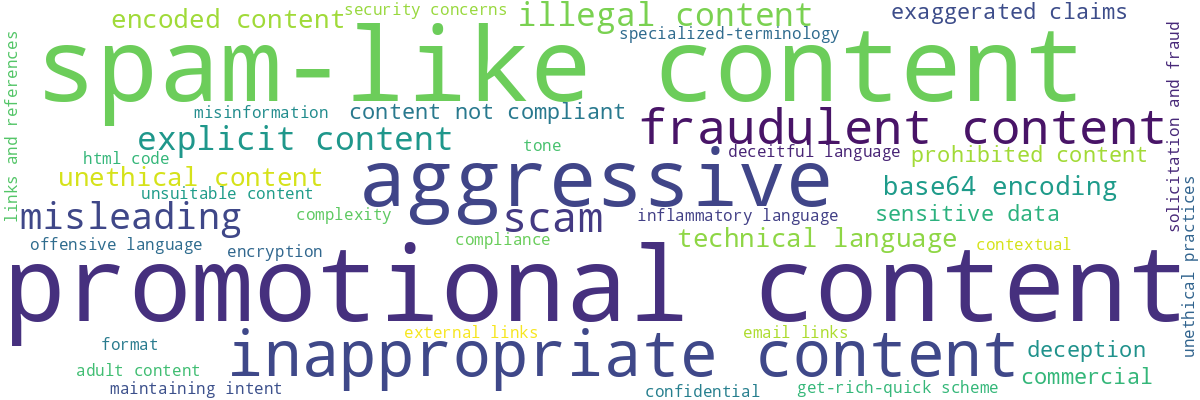}
    \caption{Word cloud with reasons given by GPT-3.5-turbo-0125 on why it did not process the API request.}
    \label{fig:wordcloud}
\end{figure}

The number of rejected emails displayed in \Cref{tbl:llm-mod} shows that 20-25\% of the requests made were rejected by GPT-3.5 Turbo.
Since the rejections affect nearly a quarter of all emails, we utilize the \texttt{failed-keyword} property of the function call to get an understanding of the most frequent reasons, displayed in \Cref{fig:wordcloud}.

After sending the same requests to GPT-4o, we observed nearly 95\% of them being blocked.
This high blocking rate was to be expected since GPT-4o operates under stricter guidelines as compared to its predecessors \cite{gpt4o:hello}.
Therefore, GPT-3.5 is the easier-to-use solution for now.

As part of our pipeline's evaluation, we compare the semantics within the email bodies before and after the modification by the LLM. 
A spam email that was rephrased and misclassified by the filter as ham is not very useful to the email's author if the original intent or message is no longer conveyed correctly.
This is also why Prompt \ref{prompt:P2} instructs the model to preserve all links.
To enable accurate comparison of text bodies, it is essential to first standardize their format.
This involves removing any line breaks and HTML tags, and reducing multiple spaces to a single space between words.
Additionally, all non-letter and non-space characters are eliminated.
Once the texts are uniformly formatted, the next step is to analyze their similarity.
This can be accomplished using GPT4All.Embed4All()\footnote{\url{https://docs.gpt4all.io/gpt4all\_python/ref.html\#gpt4all.gpt4all.GPT4All.retrieve\_model}} to obtain vector embeddings of the pre- and post-modified email bodies.
With that, we calculate the cosine similarity, quantifying the degree of semantic similarity between both texts - where $1$ indicates equal semantics and $-1$ opposite meaning.

\begin{figure}[!b]
    \centering
    \includegraphics[width=.95\textwidth]{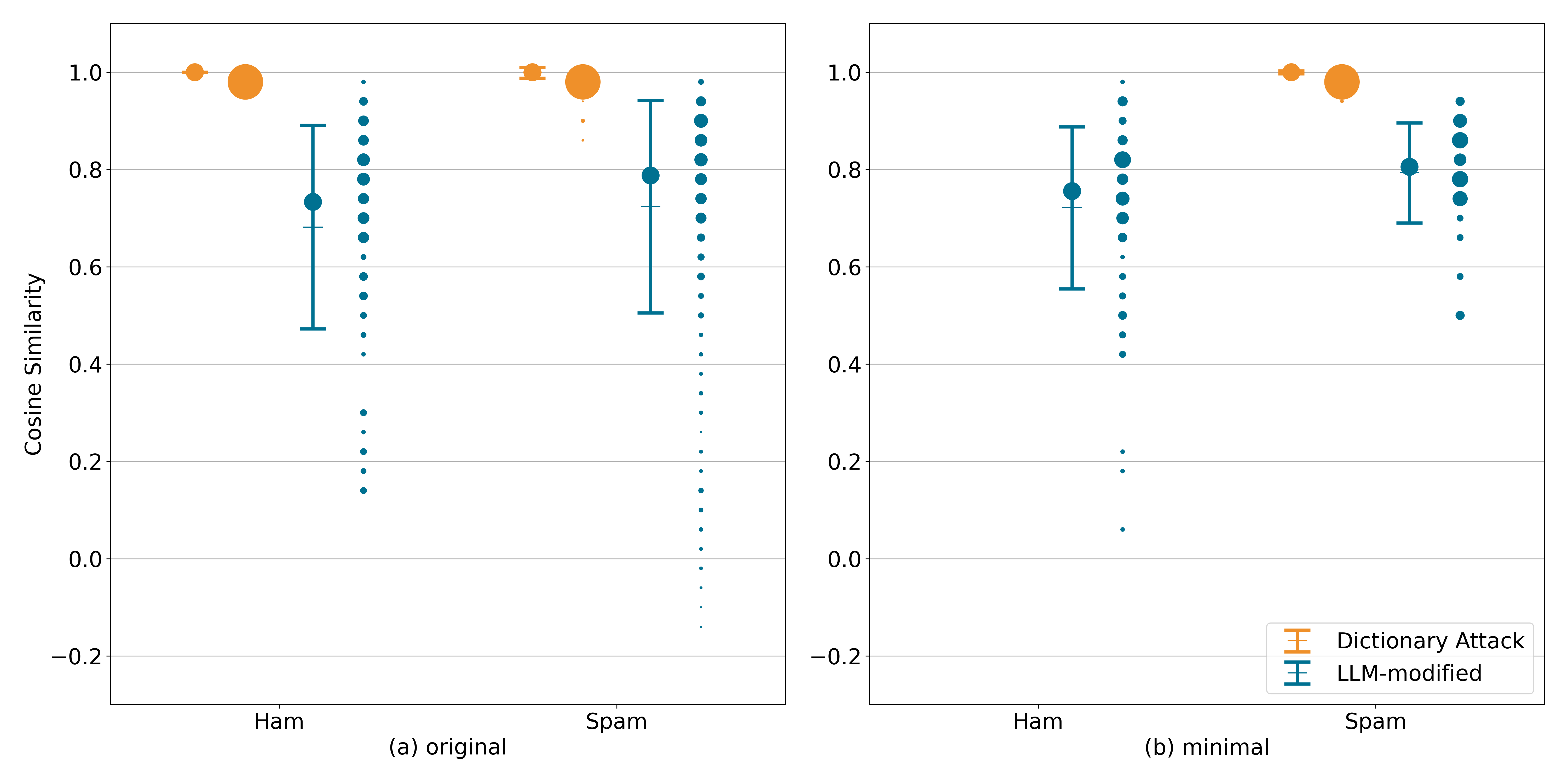}
    \caption{Cosine similarity for the datasets (a) original and (b) minimal.}
    \label{fig:cos-sim}
\end{figure}

\Cref{fig:cos-sim} displays the distribution of emails and the corresponding cosine similarities of their bodies, for both, our approach (\textit{LLM-modified}) and the low-effort approach of a dictionary-replacement attack (\textit{Dictionary Attack}).
Overall, the mean cosine similarity for our approach lies at approximately 0.8, indicating a high semantic similarity to the original text.
The variance of similarities can be explained by the fact that the LLM does not only exchange single words but entire sentences or paragraphs.
Thus, some degree of divergence from the original email bodies is expected.

To justify the need for our pipeline, we implemented a simple dictionary replacement attack, where each identified spam-like word in the email bodies was replaced with a less spam-like alternative.
To generate this dictionary, more than 750 spam-like words were extracted from \cite{mailmeteor:words} and we generated alternative, less spam-like formulations with GPT-4-Turbo.
Our findings indicate that the success rates of the dictionary replacement attack were comparatively low, with only 0.4\% for the \textit{original} and 0\% for the \textit{minimal} data set.
Furthermore, our evaluation, as shown in \Cref{fig:cos-sim}, revealed that the cosine similarities were mostly 1, indicating that the replacements, while linguistically similar, did not significantly alter the underlying message.
A result, that was anticipated, since only one word is replaced with a (semantically) similar one, not changing the underlying message too much.
These results, depicted in \Cref{tbl:success}, underscore the limitations of straightforward replacement strategies.

With a total expense of \$2.45 (approximately 0.17 cents per email) for modifying 1,443 spam email bodies, the proposed method is highly cost-effective.
Additionally, it is important to consider that each spam email is typically sent to thousands of email addresses.
This widespread distribution means the returns are significantly amplified, making the method economically advantageous.
Therefore, as LLMs are eminently accessible, relatively inexpensive, and easy to use, only requiring some prompt engineering, using LLMs was a reasonable choice.
They provide efficient and cost-effective means to process and generate large volumes of text, leveraging their accessibility and ease of use without needing extensive (local) computational resources or specialized expertise.

\section{Conclusion}\label{sec:conclusion}
In this work, we addressed the emerging threat of LLMs, with their remarkable power in generating and modifying text, to Bayesian spam filters, specifically targeting SpamAssassin and its performance against LLM-modified email contents.
We developed a pipeline that utilizes ChatGPT-3.5 Turbo to rephrase spam emails and test the spam filter's efficacy against the modified contents.
Our findings show that in our setup, SpamAssassin misclassifies 73.7\% of the LLM-modified spam emails, and after going through the entire pipeline, 95.8\% of the initial spam emails as ham.
Compared to the low-effort approach of utilizing a dictionary-replacement attack that only showed a maximum success rate of 0.4\%, our work shows much more promising results.
Combined with the GPT-3.5 Turbo's cost-efficiency (0.17ct/email), LLM-modified spam emails pose a substantial threat to Bayesian spam filters, for they successfully bypass existing filtering rules without majorly distorting the embodied message and its objective.
In our future work, we intend to evaluate the performance of our pipeline in comparison to highly configured instances of SpamAssassin and other filters.
Additionally, we plan to use various datasets and other LLMs to validate our findings.

\vspace*{4mm}
\textbf{Disclosure of Interests.} The authors have no competing interests to declare
that are relevant to the content of this article.

\bibliographystyle{splncs04}
\bibliography{references}

\end{document}